\documentclass[useAMS,usenatbib]{mn2e}
\usepackage{graphicx}
\usepackage{epsfig}
\usepackage{natbib}  
\usepackage{epstopdf}
\newcommand\araa{{ARA\&A}}%
\newcommand\apj{{ApJ}}%
\newcommand\apjl{{ApJ}}%
\newcommand\apjs{{ApJS}}%
\newcommand\aap{{A\&A}}%
\newcommand\aaps{{A\&AS}}%
\newcommand\mnras{{MNRAS}}%
\newcommand\nat{{Nature}}%
\def\simgt{\lower.5ex\hbox{$\; \buildrel > \over \sim \;$}}
\def\simlt{\lower.5ex\hbox{$\; \buildrel < \over \sim \;$}}

\def\c12{$^{12}$C}
\def\neo20{$^{20}$Ne}
\def\al27{$^{27}$Al}
\def\ne22{$^{22}$Ne}
\def\na23{$^{23}$Na}
\def\mg25{$^{25}$Mg}
\def\mag26{$^{26}$Mg}
\def\nafe{[Na/Fe]}

\def\ofe{[O/Fe]}

\def\alfe{[Al/Fe]}
\def\mgfe{[Mg/Fe]}
\def\tend{$t_{\rm end}$}
\def\taupeak{$t_{\rm acc}$}
\def\sigpri{$\tau_{\rm pri}$}

\newcommand{\msun}{\ensuremath{\, {M}_\odot}}
\newcommand{\Msun}{\ensuremath{\, {M}_\odot}}
\newcommand{\ocen}{$\omega$~Cen}

\title[Super--AGB and GC populations]{The role of super-asymptotic
  giant branch ejecta in the abundance patterns of multiple populations in globular clusters}
\author[A. D'Ercole et al.]{Annibale D'Ercole$^{1}$\thanks{E-mail: annibale.dercole@oabo.inaf.it}, Francesca  D'Antona$^{2}$, Roberta Carini$^{{2},{3}}$,  Enrico Vesperini$^{4}$, 
\newauthor
Paolo Ventura$^{2}$\\
$^{1}$  INAF- Osservatorio Astronomico di Bologna, via Ranzani 1, I-40127 BOLOGNA (Italy).\\
$^{2}$ INAF, Osservatorio Astronomico di Roma, Via Frascati 33, I-00040 Monteporzio Catone (Roma), Italy.\\
$^{3}$ Department of Physics, Universit\`a di Roma ``La Sapienza'', Roma, Italy\\
$^{4}$ Department of Physics, Drexel University, Philadelphia, PA 19104, USA
}
\begin{document}

\date{Accepted . Received ; in original form }

\pagerange{\pageref{firstpage}--\pageref{lastpage}} \pubyear{2011}

\maketitle

\label{firstpage}

\begin{abstract}
  Star formation from matter including the hot CNO processed ejecta of
  asymptotic giant branch (AGB) winds is regarded as a plausible
  scenario to account for the chemical composition of a stellar second
  generation (SG) in Globular Clusters (GCs). The chemical evolution
  models, based on this hypothesis, so far included only the yields
  available for the massive AGB stars, while the possible role of
  super--AGB ejecta was either extrapolated or not considered.  In
  this work, we explore in detail the role of super--AGB ejecta on the
  formation of the SG abundance patterns using yields recently
  calculated by Ventura and D'Antona.

  An application of the model to clusters showing extended Na--O
  anticorrelations, like NGC~2808, indicates that a SG formation
  history similar to that outlined in our previous work is required:
  formation of an Extreme population with very large helium content
  from the pure ejecta of super--AGB stars, followed by formation of
  an Intermediate population by dilution of massive AGB ejecta with
  pristine gas. The present models are able to account for the very
  O-poor Na-rich Extreme stars once deep-mixing is assumed in SG
  giants forming in a gas with helium abundance Y$>0.34$, which
  significantly reduces the atmospheric oxygen content, while
  preserving the sodium abundance.  On the other hand, for clusters
  showing a mild O--Na anticorrelation, like M~4, the use of the new
  yields broadens the range of SG formation routes leading to
  abundance patterns consistent with observations. Specifically, our
  study shows that a model in which SG stars form only from super--AGB
  ejecta promptly diluted with pristine gas can reproduce the observed
  patterns. { We briefly discuss the variety of (small) helium
    variations occurring in this model and its relevance for the
    horizontal branch morphology.}

In some of these models, the duration of the SG formation episode can
be as short as $\sim 10$ Myr; the formation time of the SG is
therefore compatible with the survival of a cooling flow in the
cluster core, previous to the explosion of the SG core collapse
supernovae.  We also explore models characterized by the formation of
multiple populations in individual bursts, each lasting no longer than
$\sim$10~Myr.
\end{abstract}


\section{Introduction}
\label{sec:intro}

The presence of multiple stellar populations, in all globular clusters
(GCs) so far spectroscopically examined, is shown with convincing
evidence in the recent analysis of \cite{carretta2009a} of about two
thousand stars in 19 GCs. Each cluster displays a sodium -- oxygen
anticorrelation (though the extension of the anticorrelation differ
from cluster to cluster) due to the existence of a population of stars
which are richer in sodium and poorer in oxygen than halo stars with
the same metallicity. The Na--O anti--correlation is typical of GCs,
whose constituent stars belong to two or more stellar populations
differing in the abundances of the elements produced by the hot CNO
cycle and by other proton--capture reactions on light nuclei.  In
fact, these chemical signatures are present also in turn--off stars
and among the subgiants \citep[e.g.][]{gratton2001,briley2002,
  briley2004}, so they can not be due to ``in situ" mixing in the
stars, but must be due to some process of self--enrichment occurring
at the first stages of the cluster life.

Photometric evidences for the presence of multiple populations are
also numerous, and sometimes suggestive of star formation occurring in
separate successive bursts. The photometric signatures of different
populations can be ascribed in part to helium differences, inferred
from the morphology of the horizontal branches (HB)
\citep{dantona2002,dc2004,lee2005}, or from the presence of multiple
main sequences (MS), in \ocen\ \citep{norris2004, piotto2005} and
NGC~2808 \citep{dantona2005,piotto2007}. The strong link between the
abundance anomalies (sodium rich and sodium poor groups) in the
cluster M~4 \citep{marino2008m4} and their location on the HB is
proven by \cite{marino2011}, who find that the blue HB stars of M~4
have high sodium, while the red HB stars have normal sodium. This
supports the interpretation that there is a slight increase in the
helium abundance of the ``anomalous" stars (the sodium rich group)
\citep{dantona2002}.

Two other important links between photometric and spectroscopic
evidence have been recently given. The first one comes from the
analysis of elemental abundances of two main sequence (MS) stars in NGC~2808
\citep{bragaglia2010} showing that the blue MS star only has
``anomalous" sodium and aluminum. This supports the interpretation
that the blue MS was formed from helium rich, CNO processed gas.  The
second link has been stressed most recently by \cite{pasquini2011}
who, directly comparing the He~I~10~830 {\AA} lines in the spectra of two
NGC~2808 giants having different O and Na abundances, find direct
evidence of a significant He line strength difference. From a detailed
chromospheric modeling, they show that the difference in the spectra
is consistent and most closely explained by an He abundance difference
between the two stars of $\Delta$Y $\geq$0.17, consistent with the
expected difference in abundance required by stellar models to account
for the blue MS.

Based on these and on a large body of other complex evidences, the
formation of globular clusters is now schematized as a two--step
process, lasting no longer than $\sim$100 Myr, during which the
nuclearly processed matter from a ``first generation" (FG) of stars
gives birth, in the cluster innermost regions, to a ``second
generation" (SG) of stars with the characteristic signature of a
distribution of element abundances fully CNO--cycled.

Two main scenarios have been proposed to identify the matter
constituting the SG stars: these could have formed either from the
gaseous ejecta of massive asymptotic giant branch (AGB) stars
\citep[``AGB scenario";][]{ventura2001, dc2004, karakas2006} or from
the ejecta of fast rotating massive stars \citep[``FRMS
scenario";][]{prantzos2006, meynet2006, decressin2007}. These studies
investigated whether the observed abundance patterns found in GCs can
be accounted for by models. Problems are present in both scenarios
\citep[see, e.g.,][for a critical review]{renzini2008}.

In order to go from a ``scenario" to a ``model", the SG formation must
be investigated quantitatively, and must rely as far as possible on
computed models. From this point of view, very few are the ``models"
available in the literature, in spite of the large number of works
recently published on this subject.  In particular, the O--Na
anticorrelation is a signature present in all cluster, but the
cluster-to-cluster differences are very large: some clusters (the most
massive ones) show very extended anticorrelations, down to values of
[O/Fe], more than 1~dex smaller than the values of the FG stars, other
clusters show a ``mild" anticorrelation, in which [O/Fe] is reduced by
$\sim$0.2--0.3~dex. In addition, \cite{carretta2009a} point out that
there is a direct correlation between the {\it minimum} [O/Fe] and the
{\it maximum} [Na/Fe] for the examined clusters. Any model, in the
end, must account quantitatively for these features.

\cite{dercole2008} followed the hydrodynamical formation of a cooling
flow at the cluster center that occurs following the epoch of
supernovae type II (SN~II) explosions \footnote{{ Actually, we
    refer here to the explosion of core-collapse supernovae, thus
    including also type Ib and Ic; however, for brevity, in the rest
    of the paper we refer to all these as SN II.}}, and the associated
loss of the remnant gas from which the FG stars had formed. The
cooling flow is due to the low--velocity stellar winds and planetary
nebulae of super--AGB and massive AGB stars, and meets the physical
conditions for a second epoch of star formation. They show that the
three populations with different helium content, necessary to explain
the presence of a triple main sequence in NGC~2808
\citep{dantona2005,piotto2007} can be reproduced if the most helium
rich population forms from the pure ejecta of super--AGB stars,
collecting in the cluster core devoid of pristine gas after the end of
the SN~II epoch. Afterwards, pristine gas is re-accreted and mixes
with the massive AGB ejecta, giving origin to the Intermediate
population having a helium content intermediate between that of the
AGB and of the pristine gas. This hydrodynamical model could
investigate only the helium content behaviour, as the yields of
super--AGB stars for elements different from helium were not available
at that time.

It is important to understand whether the AGB pollution model is
consistent with the other abundance patterns of GC stars, so
\cite{dercole2010} (hereafter D2010) developed a chemical evolution
model to test how the O--Na and Mg--Al anticorrelations could be built
up in different clusters. D2010 adopted the \cite{vd2009} yields for
massive AGB ejecta, empirically modified and extrapolated by educated
guesses to account for the super--AGB ejecta. D2010 reproduced not
only the abundance patterns of NGC~2808, but also the very different
O--Na abundance pattern of a cluster with a mild O--Na anticorrelation
like M~4. The conditions required to reproduce the mild O--Na
anticorrelations are : 1) that the Exreme population from pure AGB
ejecta could not form in this cluster (and/or in clusters with similar
patterns); 2) that the dilution with pristine gas, and the consequent
intense SG star formation due to the increased amount of available
gas, had to be delayed by $\sim$30~Myr after the end of the SN~II
epoch, a constraint that should be modeled and further explored with
appropriate dynamical simulations.

In this paper we further extend the models presented in our previous
work to include the new yields for super--AGB models from 6.5 to
8\msun\ computed by \cite{vd2011}.  In Sect.~\ref{sec:yields} we
outline the new complete set of yields now available and introduce a
hypothesis of in situ ``deep--mixing", in the high helium red giants,
that helps to deal with the extreme oxygen anomalies.  In
Sect.~\ref{sec:model} we summarize the D2010 framework and its
relevant parameters.  We present our models for the chemical evolution
of M~4 in Sect.~\ref{sec:results} and of NGC~2808 in
Sect.~\ref{sec:n2808}. Interestingly, models based on the new extended
set of yields lend further support to our previous models while
solving some of their difficulties.  In some of M~4 models the second
star formation epoch may occur in just $\sim$10~Myr, so these models
are not in contrast with the formation of SG massive stars that,
exploding as SN~II, prevent any further star formation.  We discuss in
Sect.~\ref{sec:snsg} models including an SN~II phase also for the SGs
of NGC~2808. We summarize the results and discuss the implications for
the initial mass of the FG in Sect~\ref{sec:discussion}.

 \begin{table*}
\caption{Averaged abundances in the ejecta of massive AGB and super--AGB stars.}             
\label{yields}      
\begin{tabular}{l l c l r c c c c}     
\hline       
$M/M_{\odot}$ & $\tau/10^6$$^a$ & $M_{\rm c}/M_{\odot}$$^b$ &  Y& \ofe &  \nafe &  \mgfe  &\alfe & $\log \epsilon$(Li)$^c$ \\
\hline                                                                            
3.0$^1$ & 332 & 0.76 & 0.248 &   0.92 & 1.16 &  0.57  & 0.65  & 2.77  \\  
   3.5 & 229   & 0.80 &  0.265 &    0.77 & 1.30 &  0.55  & 0.66   & 2.43 \\
   4.0 & 169.5 & 0.83 &  0.281 &    0.44 & 1.18 & 0.48  &  0.55  &  2.20\\
   4.5 & 130.3 & 0.86 &  0.310 &    0.19 & 0.97 & 0.43  &  0.85  &  2.00 \\
   5.0 & 103.8 & 0.89 &  0.324 &   -0.06 & 0.60 & 0.35  &  1.02  & 1.98 \\
   5.5 & 85.1  & 0.94 &  0.334 &   -0.35 & 0.37 &  0.28  & 1.10   & 1.93  \\
   6.0 & 71.2  & 1.00 &  0.343 &   -0.40 & 0.31 &  0.29  & 1.04   & 2.02  \\
   6.3 & 65.2  & 1.03 &  0.348 &   -0.37 & 0.30 &  0.30  & 0.99   & 2.06  \\
   6.5$^2$ & 61.5  & 1.08 &  0.352  &  -0.24  & 0.32 & 0.34& 0.91 & 2.36   \\
   7.0 & 53.7  & 1.20 &  0.358  &  -0.15  & 0.39 &    0.36  &  0.86  &  2.12  \\   
   7.5 & 46.8  & 1.27 &  0.359  &   0.01  & 0.67 &    0.415 &  0.65  & 2.75  \\
   8.0 & 38.8  & 1.36 &  0.344  &   0.20  & 1.00 &    0.440&   0.50   & 4.39 \\
\hline                                                                                                       
\end{tabular}
\leftline{$^a$Total evolutionary time until the AGB phase }
\leftline{$^b$Core mass at the beginning of the AGB phase.}
\leftline{$^c$ $\log \epsilon$(Li)= $\log (N_{\rm Li}/N_{\rm H})+12.$ }
\leftline{$^1$ Yields for 3$\leq$M/\msun$\leq$6.3 from \citet{vd2009}.}
\leftline{$^2$ Yields for 6.5$\leq$M/\msun$\leq$8.0 from \citet{vd2011} and \citet{vcd2011}}.
\leftline{$^3$ The values listed in the present Table are obtained for
  the following set of initial abundances in mass fraction: 
  H=0.75, He=0.24, } 
\leftline{Li=$10^{-11}$, C=$0.849\times10^{-4}$, N=$0.249\times 10^{-4}$, O=$0.58\times 10^{-3}$,
  Na=$0.103\times 10^{-5}$, Mg=$0.497\times 10^{-4}$, Al=$0.178\times 10^{-5}$, Fe=$0.377\times 10^{-4}$.}
\end{table*}
 
\section{Super--AGB and AGB yields}
\label{sec:yields}
\begin{figure}
\vskip -8pt
\centering{\includegraphics[width=8cm]{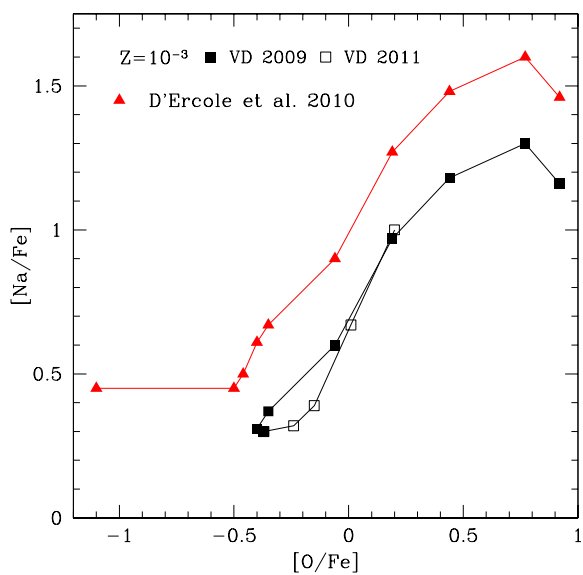}}
\caption{ The yields of metallicity Z=10$^{-3}$ from Ventura \&
  D'Antona (2009) from AGB models are plotted as full black
  squares. Masses from top to bottom are 3, 3.5, 4, 4.5, 5, 5.5, 6 and
  6.3\msun. Yields from Ventura \& D'Antona (2011) for super--AGB
  models of 6.5, 7, 7.5 and 8\msun\ are plotted as open black
  squares. Red triangles represent the yields adopted in D'Ercole et
  al. 2010: masses are 3, 3.5, 4, 4.5, 5, 5.5, 6, 6.3, 6.5 and 9\msun.}
\label{f1} 
\end{figure}

In Table~\ref{yields} we list the average abundances in the ejecta of
AGB and super--AGB stars adopted in this work. These yields have been
calculated by \cite{vd2009} (see their Table 2) and further extended
with the calculations of the yields for super--AGB stars by
\cite{vd2011} and \citet{vcd2011}. The latter works show the
continuity between the two set of yields. The models have been
computed by using the same input physics and nuclear reaction rates,
so the results can be matched together. We refer to \cite{vd2011} also
for a comparison with other super--AGB results in the literature
\citep[notably by][]{siess2010}. Fig.~\ref{f1} shows the O--Na yields
for these models, together with the yields adopted in D2010.

Notice that the maximum mass that does not ignite as SN~II in the
super--AGB models here adopted is 8\msun. Thus the second generation
formation in these new chemical evolution models can not start before
$\simlt$40Myr from the FG birth, as this is the evolutionary time
computed for the 8\msun\ star. In D2010 we assumed that super--AGB
would include the 9\msun\ star, evolving at $\sim$32~Myr. The minimum
mass leading to SN~II explosion (and thus the duration of the SN~II
epoch) depends mostly on the assumptions made on the core overshooting
during the core hydrogen burning. Our models employ a moderately large
overshooting \citep[e.g.][]{schaller1992,ventura1998a}. If this is
reduced, the minimum mass of possible SN~II progenitors increases, and
the lifetime of the SN~II epoch decreases.  This input defines not
only the beginning of the phase during which the low velocity winds
accumulate for the SG formation (after the FG SN II stage), but also
the duration of a possible, subsequent, SG SN~II phase (from $\sim$30
to $\sim$40~Myr, see Sect.~\ref{sec:snsg}).

\subsection{Sodium}
\label{sodium}
By increasing the initial mass from 3\Msun\ to 8\Msun, the Na and O
values provided by the models ejecta first decrease, reach a minimum
value in the yields (at masses $\sim$6.5\msun), and then they both
revert back toward higher values. Notice that the location in the
Na--O plane of the yields of super--AGBs for the masses
6.5$\leq$M/\msun$\leq$8 is not different from that of stars with lower
masses, it is simply reversed.  This behaviour is due to the huge mass
loss rates of super--AGBs \citep{vd2011} and differs from the values
adopted in D2010 in which the last point in the yield table was taken
to be at very low oxygen ([O/Fe]=--1) and moderate sodium.

The comparison between yields and the data for single-metallicity GCs
clearly shows that to reproduce the observed Na--O anticorrelation it
is necessary to dilute the AGB ejecta with pristine gas
\citep{bekki2007,dv2007,dercole2008, dercole2010}.

Although the yields of massive stars do not show the strong direct
correlation of Na and O, but a moderate anticorrelation, dilution is
also needed in models based on the FRMS ejecta
\citep{prantzos2006,decressin2007,lind2011}, and seems to be present
in all clusters. Detailed hydrodynamical models able to explain the
dynamics and the presence of pristine gas during the SG formation are
still lacking. The issue of dilution is discussed in detail by
\cite{dercole2011}.
 
The Na abundances adopted in D2010 are +0.3~dex larger than the values
reported in \cite{vd2009} and adopted in the present work. This
assumption was necessary to succesfully reproduce the observed
anticorrelations, and was theoretically justified by assuming a 50\%
increase in the $^{20}$Ne(p,$\gamma$) reaction rate and a larger
$^{20}$Ne abundance in the initial gas mixture. In the present work,
use of super--AGB yields makes this adjustment necessary only for the
specific case of M 4 in which a $^{20}$Ne overabundance is actually
present (see Sect.~\ref{sec:model}).

\subsection{Oxygen}
\label {O16}
The computed super--AGB models show that the sodium and oxygen yields
increase with the mass, and that [O/Fe] never reaches values smaller
than $-0.4$ (cf.  Table~\ref{yields} and Fig.~\ref{f1}). This
behaviour raises a question on the origin of the very low oxygen
abundances ([O/Fe]$\sim -1$) found in the giants of some clusters
(these are the stars classified as Extreme second generation by
Carretta et al. 2009a).  We propose to solve the issue with a single
additional assumption.  We attribute the very small oxygen abundances
to an ``in situ'' deep--mixing, acting only {\it in the progenitors
  having a very high helium content}. In these stars, the small
molecular weight discontinuity during the red giant branch evolution
may not be able to preclude deep-mixing and lowering of oxygen in the
envelope. Stars having the largest helium abundances would then
signficantly reduce their atmospheric oxygen content, while preserving
a similar sodium abundance \citep{dv2007}\footnote{Historically,
  deep-mixing was modeled in giants starting their life from a
  standard abundance mixture, and produces a sodium increase
  concomitant to the oxygen depletion, dealing with regions in which
  neon has been processed to sodium by proton captures. These models
  were in reasonable agreement with the whole sodium versus oxygen
  anticorrelation (see, e.g. Weiss, Denissenkov \& Charbonnel 2000,
  their fig. 9), but the deep-mixing interpretation had to be
  abandoned when the chemical anomalies were shown to be present in
  turnoff stars. If deep-mixing is instead allowed by the high helium
  content of the Extreme SG component, we must take into account that
  these stars are formed from AGB processed matter, in which the
  initial neon has already been converted to sodium. In this case, no
  additional sodium is produced inside the star during the hydrogen
  burning stages, as shown by \cite{dv2007}.}. The presence of [O/Fe]
abundances smaller than those predicted by AGB models only among the
giants, and not among the turnoff or subgiant stars so far examined
\citep{carretta2006}, and in stars belonging to clusters also showing
the signature of a very helium rich MS, such as NGC~2808 and \ocen,
are a possible argument in favour of this interpretation. Obviously,
the determination of a very low oxygen content ---similar to the one
in the most anomalous red giants--- in the atmosphere of blue MS stars
in these clusters would falsify this suggestion \citep[][were not able
to measure oxygen in their spectrum of the blue MS in
NGC~2808.]{bragaglia2010}.

We apply the following deep-mixing scheme to our models: if a
population of SG stars is born from the pure ejecta of super--AGB
stars and massive AGBs, in which the helium content is Y$>$0.34, we
assume that the surface abundances in the atmospheres of the SG giants
is due to deep-mixing; as explained above, this has the effect of
reducing the oxygen abundance while leaving the Na abundance
unaltered.

The following steps are followed: 1) we examine the helium content in
the star--forming gas at each time step; this helium content is the
average content of the gas reservoir, either due to the super--AGB or
AGB ejecta, in absence of pristine gas, or averaged with the
re--accreted pristine gas; 2) if the helium content in the star
forming gas is Y$>$0.34, we adopt a ``deep--mixing" scheme to provide
the oxygen content {\it at the surface} of giants.  In the absence of
non parametric models, the relation chosen to fix the oxygen content
after the deep-mixing has some degree of arbitrariness. We decided to
link the value to the helium abundance of the gas forming the
stars. [O/Fe] is linearly interpolated between the values --0.4 and
--0.9, assumed respectively for helium content Y=0.34 and
Y=0.36. Consequently, when the ejecta of the mass 7.5\msun\ form
stars, they will reach the minimum oxygen values when evolving as
giants.  We assume: [O/Fe]=--0.4--(Y--0.343)$\times 37.5$. We point
out that this expression is adopted as an example only, and other
--arbitrary-- formulations could have been chosen.  3) if the average
helium abundance is smaller than the values for which deep-mixing
occurs, the [O/Fe] of the SG stars is equal to the average between the
standard [O/Fe]$\sim$0.4 of the pristine matter, and the mildly
depleted [O/Fe] of Table~\ref{yields}.

This unique assumption on the simulations will change dramatically our
understanding of the O--Na anticorrelation in clusters that do not
show extreme anomalies.

\subsection{Magnesium and Aluminum}
As for magnesium and aluminum , in this paper we adopt the yields
obtained in our standard models \citep{vd2009} in which the NACRE
upper limits for the aluminum production by proton capture on \mg25\
and \mag26 \citep{angulo} are used. The super--AGB computations have
shown that the maximum Mg depletion is achieved in models of
5--6\msun, and not in the super--AGB models, contrary to the
extrapolation suggested in D2010. The extent of possible variation of
Mg and Al yields by varying the input physics is discussed in
\cite{vcd2011}.

\subsection{Helium}
As discussed in the literature \citep{vdm2002}, the helium abundance
in the ejecta of { massive} AGB stars is mainly due to the
second--dredge up, as the thermal pulse phase in the mass range of
interest for the formation of the second generation is so short that
the episodes of third dredge up do not increase significantly the
value achieved at the second dredge up. { The same happens in
  super--AGB stars, whose evolution is faster than that of massive
  AGBs and whose thermal pulses are less energetic and less effective
  for what concerns the third dredge up. Therefore, the helium yield
  of super--AGB stars, contrary to what happens for the other chemical
  elements of interest here, does not depend on the assumptions made
  for the mass loss rate.}  The average helium abundance increases
steadily with the mass, reaching a maximum value of mass fraction
Y=0.359 for the 7.5\msun model. In the 8\Msun\ evolution, the total
helium in the ejecta is down to Y=0.344. As discussed in
\cite{vd2011}, this is due to the different behaviour of this model,
that ignites carbon in the core {\it before} the second dredge
up. Although the results of the 8\msun\ evolution must be taken with
caution for what concerns the other elemental abundances, that
dramatically depend on the huge mass loss suffered by this model
\citep[following the choice made in ][for the formulation of the mass
loss rate]{vd2011}, it is difficult to think that a decrease of the
second dredge up effects is physically motivated. Therefore, these
models can not explain the Extreme multiple generations of GCs, if we
have to take at face value the helium contents Y$>$0.38 that compete
to the blue main sequences \citep{norris2004, piotto2007} or even the
value Y$\sim$0.42 attributed to the blue hook stars in NGC~2419 by
\cite{dicriscienzo2011}. We follow this line of reasoning: we must not
take at face value neither the fit of observations with models
\citep[see, e.g., the alternative interpretation by][on the Y values
to be attributed to the helium rich main sequence]{portinari2010} nor
the mediated proposals of precise very large values of Y from the
analysis of the horizontal branch stars, as often pointed out by the
same authors \citep{dantonacaloi2008}, but only the existence, in
these clusters, of well defined helium rich sequences. In the absence
of direct measurements of the helium abundance, we will assume that
the average value of Y of our super--AGB models is more than adequate
to describe the Extreme populations of these clusters. Notice that
\cite{siess2010}, in spite of using different physical inputs, reaches
a maximum value (Y=0.375) in his super--AGB models of the same
metallicity. { Other models in the literature
  \citep[e.g.][]{bs1999, doherty2010, lagarde2011} reach values of
  helium abundance in the same range (Y$\sim$0.33 -- 0.38), in spite
  of differences in codes, numerical inputs and physical approaches in
  the mixing formulation (see for discussion Ventura 2010 and
  D'Antona, Charbonnel et al. 2012, in preparation).  }
\begin{table*}
\caption{Dynamical input parameters for the simulations}             
\label{t2}      
\begin{tabular}{l c c c c c c c c c cccccc }     
\hline       
Cluster & Model & SG formation & \taupeak$^{(a)}$  & \sigpri & \tend & $(t_{\rm f}$) $\delta t$$^{(b)}$   &$\rho_{\rm *,FG}$  &$\rho_{\rm 0,pr}$ & $\nu$ & $x$ \\
&&& (Myr) & (Myr) & (Myr)& (Myr) &\msun pc$^{-3}$ &&\\
\hline                                                                            
M~4             &   0    & continuous  &  65  &  2   &   105  & & 9.4 &0.091 & 0.1 & 0.7  \\ 
M~4             &   1    & continuous  &  43  &  1   &   100  & & 60 &0.085 & 0.1 & 0.5 \\
M~4             &   2   & continuous  &  43  &  2  &   60  & & 940 &0.05 & 0.1 & 0.5 \\
M~4             &   3   & accum.  &  43  &  2  &   50  &  (45) &940 &0.04  &0.5 & 0.5 \\
M~4             &   4   & accum.  &  43  &  2  &   55  & (45) &940 &0.04 & 0.5 & 0.5  \\
M~4             &   5   & accum.  &  48  &  10  &   58  & (48) & 940 &0.08 & 0.5 & 0.5 \\
NGC 2808  &   0  & continuous   &   58  &  6   &   68  & & 240 & 0.01 &1  & 0.5   \\
NGC 2808  &   1  & continuous   &   65  &  8   &   90  & & 240 &0.0095&1  & 0.4 \\
NGC 2808  &   2  & accum.   &   65  &  8   &   90  & (50) & 240& 0.0095&  1 & 0.4 \\
NGC 2808  &   3  & bursts  &   87  &  10   &   100  & 50-80&240 &0.0045& 1&0.4  \\
NGC 2808  &   4  & bursts  &   87  &  10  &   90  & 50-80& 240 & 0.0045 & 1 & 0.4  \\
NGC 2808  &   5  & bursts   &   87  &  10   &   90  &50-80& 240 & 0.02 & 1 &  0.4   \\
\hline                                                                                                       
\end{tabular}
\leftline{$^{(a)}$ times in Myr.}
\leftline{$^{(b)}$ $t_{\rm f}$: time until which the AGB ejecta are accumulated before star formation begins} 
\leftline{$^{(b)}$ in models N2808--3, 4 and 5 we list $\delta t$, the time interval during which there is no star formation.}

\end{table*}

\section{The chemical evolution model}
\label{sec:model}
The chemical evolution is computed following equations 2--5 of
D2010. According to the framework presented in D2010, FG stars are
already in place and have the same chemical abundances of the pristine
gas from which they form; the SG stars form from super--AGB and AGB
ejecta gathering in the cluster centre through a cooling flow and
partially diluted with pristine gas accreting on the cluster core with
(possibly) a time delay with respect to the beginning of the SG
formation epoch.

In order to preserve the cooling flow, we still assume that only SG
stars that do not explode as SN~II can form
\citep{dercole2008}.\footnote{The following will show that this
  limitation in the IMF can be relaxed for clusters showing a mild
  O--Na anticorrelation, see Sect.~\ref{sec:snsg}.}  Another reason to
exclude the SG SN~II is that normally GCs show a very small spread of
iron \citep{carretta2009ferro}, indicating that the SG stars hardly
have been polluted by SN~II ejecta (neither of the FG nor of the SG).
The parameters characterizing the model are the same ones listed in
D2010:
\begin{enumerate}
\item \taupeak : time at which the maximum accretion of pristine matter occurs;
\item \sigpri : timescale of the pristine gas accretion process;
\item \tend: time at which the SG star formation ends;
\item $\rho_{\rm{*,FG}}$: densiy of FG stars;
\item  $\rho_{\rm{0,pr}}$: densiy  regulating the amount of available pristine gas, normalized to $\rho_{\rm *,FG}$;
\item $\nu$:   star formation efficiency
\item $x$ = $\rho_{{\rm *,SG}}/\rho_{{\rm *tot}}$: ratio between the SG stars and the total nowadays alive stars ($\rho_{{\rm *,tot}} = \rho_{{\rm *,FG}} + \rho_{{\rm *,SG}}$)
\end{enumerate}

An exploration of this whole set of parameters has been presented in
D2010. In the following we will vary only the most critical parameters
(\taupeak, \sigpri, \tend, $\nu$\ and $\rho_{\rm 0,pr}$).  Notice that
\tend\ is not necessarily linked to \taupeak\ and \sigpri, as the
phenomenon causing the end of the SG star formation (e.g. the phase of
type I supernova explosions, cf. D2010) is not related. The ratio $x$,
in most GCs, is larger than 0.5
\citep{dantonacaloi2008,carretta2009a}. This current value depends not
only on the duration and efficiency of the star formation of the SG,
but also on the subsequent loss of FG stars \citep[see,
e.g. ][]{dercole2008}. Although the loss of FG stars does not enter
directly into the models (it is in fact absent in equations 2--5 of
D2010), its value can be inferred a posteriori from the fit of
specific data for the clusters. In the examples of
Sect.~\ref{sec:results} we adopt $x=0.4$ or 0.5 for NGC~2808 and
$x=0.5$ or 0.7 for M~4.
 
The initial values for the pristine gas abundances can be inferred
from the observed data. For example, the initial Mg and Al abundances
in the cluster M~4 must be consistent with the observed values
[Mg/Fe]=0.5 and [Al/Fe]=0.5, while we adopt more standard values of
[Mg/Fe]=0.3 and [Al/Fe]=0 for the initial abundances in NGC~2808.  M4
FG stars also exhibit a small increase ($\sim +0.05$dex) in the
initial oxygen content.  { This calls for a variation of the Mg and
  Al abundances of the AGB ejecta given in Table \ref{yields}, as
  these latter are computed for the initial values [Mg/Fe]=0.4 and
  [Al/Fe]=0. Magnesium and Aluminum were modified on
  the basis of the details of the various channels of the Mg-Al
  nucleosynthesis, discussed in \citet{vcd2011}. The change in the
  oxygen yields is simpler, because this element is destroyed with
  continuity during the whole AGB phase, which allows a straight
  scaling relation. }  Further, the overabundance of
$\alpha$-elements in M~4 should also imply an overabundance of
neon. Adopting the relation discussed in D2010 for the sodium
production by AGB stars ($\delta$ [Na/Fe]$\simeq
\delta$[$^{20}$Ne/Fe]), we increased the sodium yield values of
Table \ref{yields} only for the specific case of M~4. { Table \ref{t3}
summarizes all the variations adopted in our models.}

\begin{table*}
\caption{Chemical input parameters for the simulations}             
\label{t3}      
\begin{tabular}{l c c c c c c c c }     
\hline       
Cluster & $\delta$[O/Fe] &$\delta$[Na/Fe] & $\delta$[Mg/Fe] & $\delta$[Al/Fe] & O$_{\rm in}$ & Mg$_{\rm in}$ &  Al$_{\rm in}$ \\
& $^{(a)}$&$^{(a)}$&$^{(a)}$&$^{(a)}$& $^{(b)}$&$^{(b)}$& $^{(b)}$\\
\hline                                                                            
M~4            &0.05 & +0.2$^1$  & +0.03 &+0.05 &580&60& 6 \\ 
NGC 2808 & 0.0 &0.0 & -0.05 & -0.10 & 580 & 40 & 2.5  \\
\hline                                                                                                       
\end{tabular}
\leftline{$^{(1)}$ for all the M 4 models but M 4-0, for which $\delta$[Na/Fe]=0.3.}
\leftline{$^{(a)}$ difference in the AGB yields with respect to Table 1.}
\leftline{$^{(b)}$ mass fraction of the element in the pristine gas, in units of 10$^{-6}$.}
\end{table*}

\section{Results: M~4}
\label{sec:results}
\subsection{Continuous star formation}
\label{subsec:cont}
\begin{figure}
\vskip -25pt
\centering{\includegraphics[width=8cm]{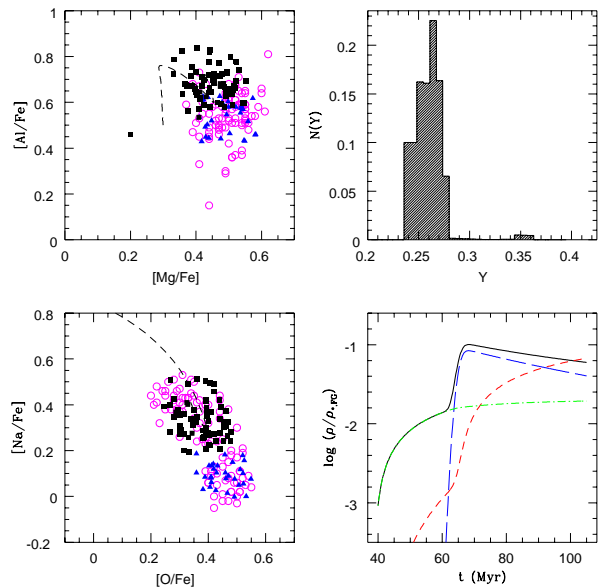}}
\vskip -50pt 
 \caption{Simulation for the SG of M~4, following the criteria of
   D2010 (case M~4-0).  The top-left panel and the bottom-left panel
   show the [Mg/Fe]-[Al/Fe] and the [O/Fe]-[Na/Fe] diagrams,
   respectively. Data (open circles) are from Marino et al. (2008).
   Squares and triangles represent a sampling of the SG and FG stars
   by our model, respectively.The dashed lines represent the gas
   trajectory within the diagrams; the sampled stars in principle
   would be located on this line, but we introduced a random scatter
   in the range 0.1 dex in their coordinates in order to mimic the
   observational errors.  The top-right panel illustrates the stellar
   He distribution. In the bottom-right panel the evolution of the
   following quantities is reported: total amount of gas (solid line),
   stellar ejecta (dot-dashed line), pristine gas (long-dashed line),
   SG stars (dashed line).  }
\label{f4} 
\end{figure}

In D2010 we presented a model to fit the M~4 O--Na and Mg--Al data by
\cite{marino2008m4}. In Fig.~\ref{f4} we reproduce a model with the
same dynamics (M~4-0 in Table~\ref{t2}) but with the abundances
calculated using the new Table~\ref{yields} (and Table \ref{t3}). The bottom right panel
shows the role of ejecta (dash--dotted line) and of the pristine gas
accretion (long--dashed line) and their timing in the detailed
formation of the SG (dashed line). The bulk of the SG formation occurs
when the pristine matter is re-accreted, at $\sim$65 Myr, and lasts
for $\sim$40 Myr more.  The dashed lines in the left panels show the
path of the O--Na and Mg--Al abundances along the simulation. We see
that, when dilution with pristine matter sets in, the values tend
towards the FG abundances. As we discussed in D2010, this choice of
input parameters provides agreement with the data if the sodium
abundance is assumed to be larger by +0.1~dex, with respect to the
yields adopted in that work, that is, larger by +0.3~dex with respect
to Table~\ref{yields}.  As discussed in Sect.~\ref{sodium}, this would
mean to assume that the neon abundance in the gas forming the FG stars
of M~4 is +0.3~dex larger than in other cluster--forming regions, in
particular in NGC~2808, for which we used directly the values in
Table~\ref{sec:yields}. The top right panel of Fig.~\ref{f4} shows the
helium distribution. The total spread is $\delta$Y$\sim$0.04, and
could be revealed by a careful analysis of the MS.

\begin{figure}
\vskip -25pt
\centering{\includegraphics[width=8cm]{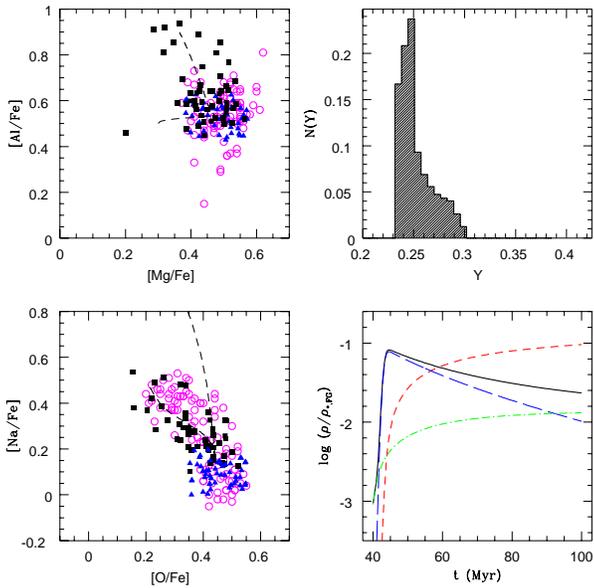}}
\vskip -50pt
\caption{Model M~4-1. In this model the pristine gas accretion is
  greately anticipated with respect to the model M 4-0 of
  Fig. \ref{f4} (symbols as in Fig. \ref{f4}).}
\label{f4a} 
\end{figure}

The new super--AGB yields broaden the range of possible star formation
routes that can satisfactorily fit the M~4 data, showing that
different re-accretion histories may lead to similar mild O--Na
anticorrelations found in numerous GCs.  In the new models star
formation can begin much earlier: a first example is shown in
Fig.~\ref{f4a} (case M~4-1). The star formation efficiency in this
model has still the low value $\nu$=0.1 assumed in the previous case,
and no star forms directly from super--AGB ejecta (as we see from the
distribution of the helium abundance).  Here (and in all the following
models of M 4) the adopted excess of sodium abundance is reduced to
$\delta\rm{[Na/Fe]}=0.2$, and we assume $x$=0.5 instead of 0.7
considered in the previous model for the sake of comparison with the model of
D2010. A fraction of 50\% of FG stars is in fact consistent with the
fraction of red HB stars given in \cite{marino2008m4}. Recent analysis
of UV based colors of the red giants actually show that the RG branch
is splitted into a Na--rich (redder) and Na--normal (bluer) part, in
balanced numbers (Milone 2011, private communication).

Model M 4-1 requires the presence of pristine gas since the early
phases of SG formation to efficiently dilute the super--AGB matter.
Thus the {\it strong dilution of the high--helium super--AGB ejecta}
leads to helium abundances not much larger than the pristine helium
abundance in the gas forming the SG.  As discussed in Sect.\ref{O16},
the oxygen content is not reduced by deep-mixing: the SG ends up
forming a group of stars with relatively large sodium (also thanks to
the inclusion of $\delta$[Na/Fe]=+0.2 in the yield table, as discussed
above), but with oxygen depleted only by $\sim$0.1~dex.  The small
helium spread shown by this simulation is probably undetectable on the
main sequence, but can lead to a distribution of the HB stars such
that the Na--rich stars are bluer than the Na--normal stars, as found
by \cite{marino2011}.

\begin{figure}
\vskip -25pt
    \centering{\includegraphics[width=8cm]{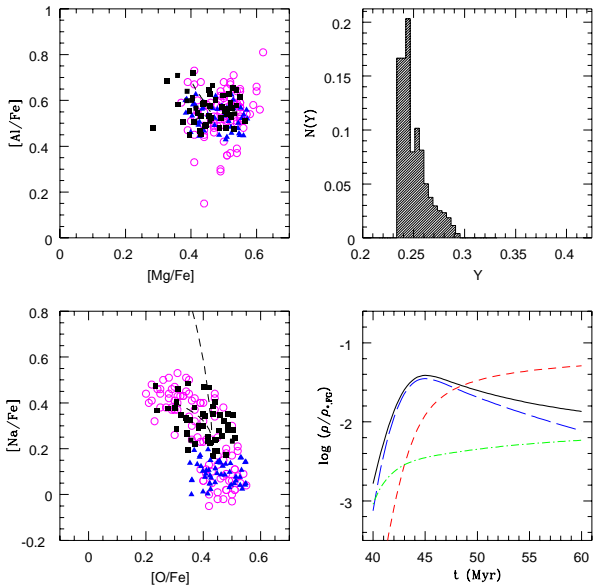}}
\vskip -50pt
\caption{Model M~4-2. The strong, early dilution or super--AGB ejecta
  with pristine gas allows the formation of the SG at early times aftr
  the end of the SN~II epoch (symbols as in Fig. \ref{f4}). The time
  evolution is much shorter compared to models M 4-0 and M 4-1.}
\label{f5} 
\end{figure}

We also run the model M 4-2 (Fig. \ref{f5}) to illustrate how
acceptable solutions can be obtained even for significantly shorter
evolutionary times. This shows that this kind of models are flexible
enough to reproduce the observations in a variety of conditions; this
also means, however, that these models can provide only limited
constraints on the gas hydrodynamics and the related star formation
history.

\begin{figure}
\vskip -33pt
\centering{\includegraphics[width=8cm]{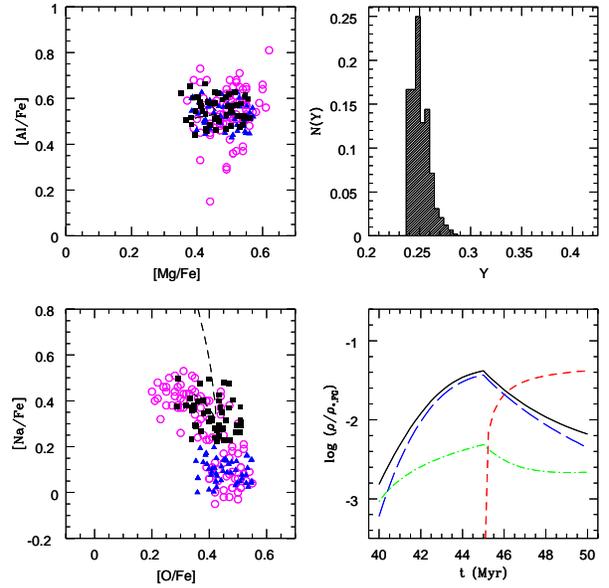}}
\vskip -50pt
\caption{Model M~4-3. The pristine gas and super--AGB ejecta
  accumulate and a short--lasting burst of star formation occurs at
  45~Myr (symbols as in Fig. \ref{f4}).}
\label{f6} 
\end{figure}

\begin{figure}
\vskip -25pt
\centering{\includegraphics[width=8cm]{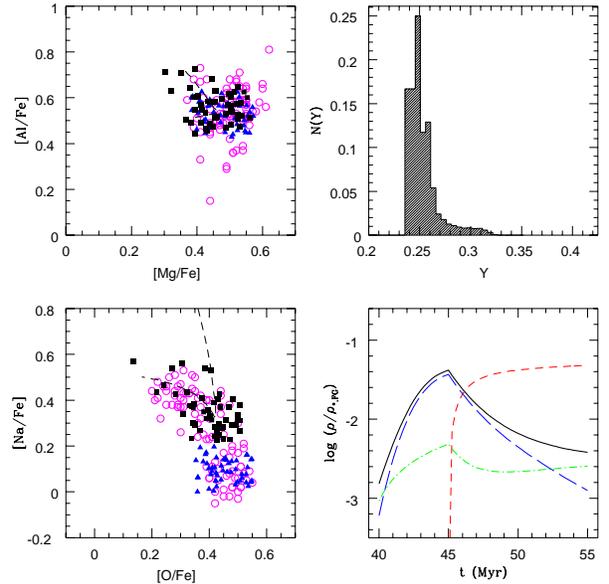}}
\vskip -50pt
\caption{Model M~4-4. The same input parameters of model M~4-3, but
  the evolution lasts for 5~Myr longer, when the pristine gas is
  already exhausted, and a tail of helium--rich stars is formed from
  pure AGB ejecta (symbols as in Fig. \ref{f4}).}
\label{f7} 
\end{figure}

\begin{figure}
\vskip -25pt
\centering{\includegraphics[width=8cm]{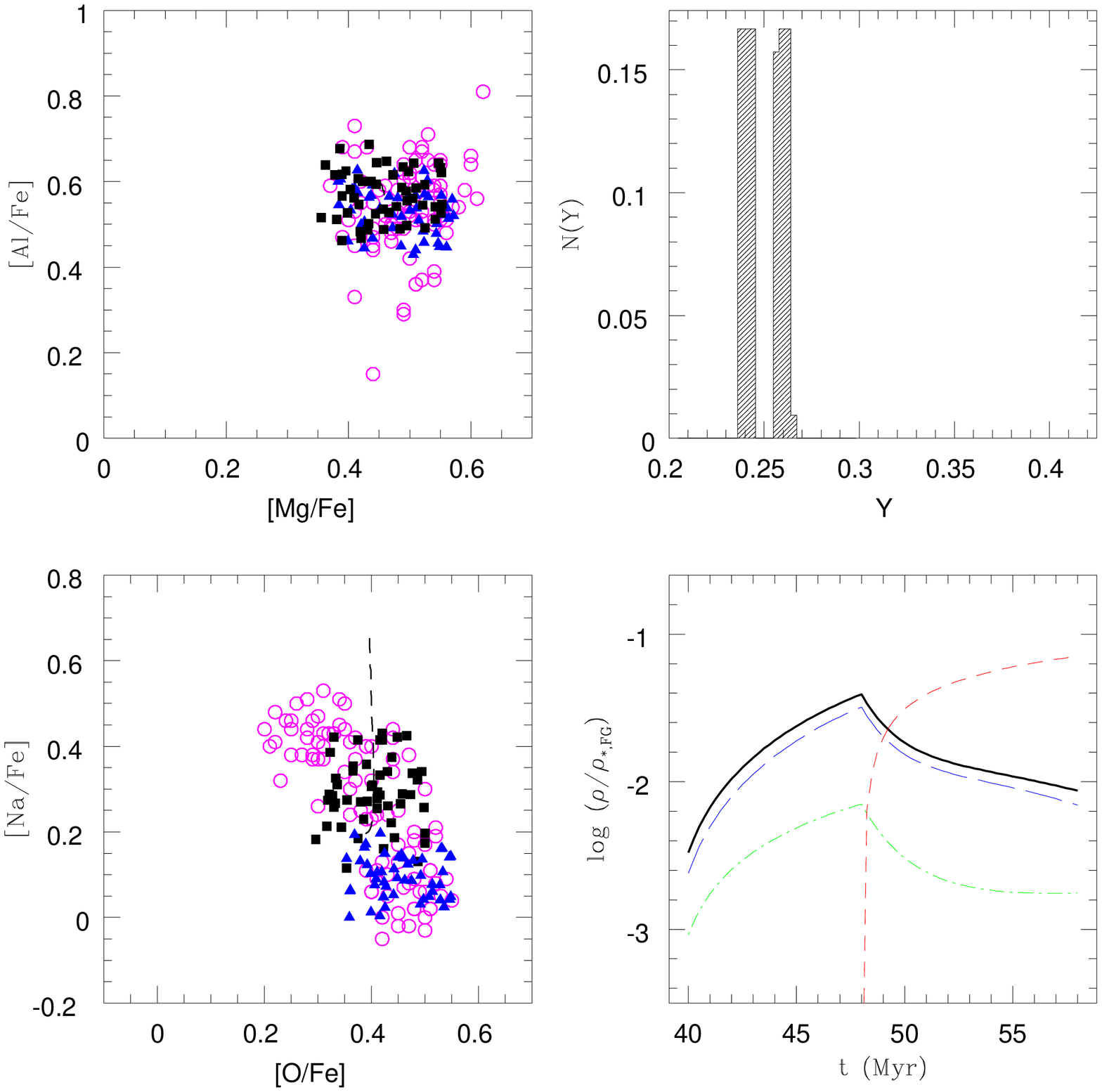}}
\vskip -50pt
\caption{Model M~4-5. The burst of star formation occurs after
  accumulation until an age of 48~Myr: FG and SG stars are born with a
  finite, although very small, difference in helium content (symbols
  as in Fig. \ref{f4}).}
\label{f8} 
\end{figure}

\subsection{Star formation after gas accumulation}
\label{subsec:accum}
In Fig.~\ref{f6} we model a different case (M~4-3). Here we adopt
$\nu$=0.5, and let the super--AGB ejecta and pristine gas accumulate
up to an age of 45~Myr, when a burst of formation occurs, lasting
until a total age \tend=50~Myr.  The process of SG formation must stop
very soon with these parameters.

In Fig.\ref{f7} (case M~4-4), we assume \tend=55~Myr, and a small tail
of stars with larger helium content (and lower [O/Fe]) is formed from
pure AGB ejecta in the time interval 50--55~Myr; such a tail is not
present in the data, but it is possible to increase \sigpri\ to avoid
this problem. Our aim here is to show how much a small variation of
the input parameters may lead to different results, {\it when SG stars
  form from super--AGB ejecta}, i.e. when, contrary to model M 4-0,
shorter evolutive times are involved.

To further illustrate this point, we show another simulation in
Fig.~\ref{f8} (Case M~4-5), with a longer timescale for the SG
formation. Interesting to notice, the helium content of the SG is
again only slightly increased, but here it is well separated from the
FG value. Possibly, a careful color analysis of the main sequence
width following the lines described, e.g., by \cite{milone2011a} could
reveal this bimodality in helium abundance, and discriminate between
this model and the previous one. Models similar to this one may be
able to reproduce the abundances in other clusters showing a bimodal
HB.

\subsection{Which model is more adequate?}
Thus, there are two main different modalities in which the AGB
pollution may lead to a viable model for M~4: either the re-accretion
of pristine gas is delayed by more than 20~Myr (model M~4-0 and
D2010), or it occurs early after the SN~II epoch (models M~4-2 and
following). The choice between these possibilities can rely upon the
distribution of other elemental abundances in the FG and SG. In
particular, the model M~4-0 shows a small increase in the aluminum
abundance of the SG (Fig.~\ref{f4}, top left
panel). \cite{marino2008m4} data, and also the recent data by
\cite{villanova2011}, do not show a clear aluminum trend.
In addition, the D2010 and M~4-0 models show a finite spread in
the helium abundance of the two populations, and model M~4-5 a
bimodality in the helium content.

{ The helium abundance may offer another possible way to choose
  among the models, by examining the morphology of the HB, and the
  abundances of sodium and oxygen at different locations along the HB.
  Originally, \cite{dantona2002} proposed to explain the color spread
  in the HB of most GCs not as an effect of a mass loss spread, as
  generally thought, but as an effect of helium increase from the
  cooler to the hotter side. At a given age, the mass of stars
  evolving off the main sequence is smaller for larger helium
  abundances. This implies that, for a similar mass loss, stars with a
  larger helium abundance end up with a smaller mass and thus at a
  bluer location on the HB. Depending on the metallicity, a very small
  helium increase (from Y=0.24 to Y$\simeq$0.27) is necessary to
  produce the gap ---a real lack of stars in the RR~Lyr region---
  between the red clump and the blue HB in NGC~2808 \citep{dc2004}. In
  clusters like M~4, having a more continuous distribution between red
  and blue, with the presence of several RR~Lyr stars, the helium
  variations may have been less abrupt. Marino et al. (2011) found
  that the blue HB stars of M~4 have high Na, while the red HB stars
  have normal Na. Furthermore, \cite{villanova2012} have recently
  observed six blue HB stars in M~4, finding that they have high Na,
  low O and helium content Y$\sim$0.29.  A similar location of
  Na--rich, O--poor stars at the blue side of the HB has been recently
  discovered by Gratton et al. (2011) in NGC~2808, confirming the
  scenario. We remark that this same analysis shows {\it a mild
    anticorrelation of the color with the Na/O ratio within the red
    clump stars,} so that, contrary to the simple interpretation, even
  the red side of the HB may be not fully populated by FG stars, but
  also by a few SG stars having a helium content barely larger than
  the FG value. Also in NGC~1851 Gratton et al. (2012) have found a
  few Na-rich stars at the hotter end of the red clump. These cases
  show that extreme care is necessary before drawing firm conclusions
  on the number ratio of FG to SG stars from simplified analyses. }.

\begin{figure}
\vskip -40pt
\centering{\includegraphics[width=8.2cm]{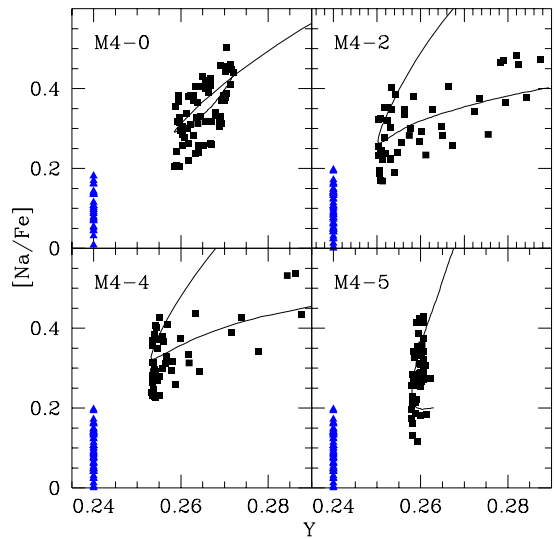}}
\vskip -50pt
\caption{ Sodium versus helium abundance in the labelled
  simulations. The line represents the Y--Na temporal path of the
  model yields during the formation of the SG. As for other figures,
  the number of points along different portions of the lines is
  proportional to the star formation rate (symbols as in
  Fig. \ref{f4}).}
\label{figuranuova} 
\end{figure}

Figure \ref{figuranuova} illustrates the different ways in which
helium is correlated to sodium in our simulations.  While in model
M4-0 the difference in helium between SG and FG stars is $\delta$Y$
>~0.02$, other models have a population of SG stars at Y$\sim 0.25$
(model M4-2) or 0.255 (model M4-4).  Model M~4--5 shows that it is
possible to obtain a mono--helium SG (in our case Y$\sim$0.26) with a
large sodium spread. From these examples, it appears that the models
employing the super--AGB may be able to explain the presence of SG
stars on the blue side of the red clump of NGC~1851\footnote{ We
  must keep in mind that the case of NGC1851, however, is more complex
  than those we are exploring here. In this cluster, metallicity and
  s-process element differences are also present and a significant age
  difference between the two populations is necessary to explain the
  complex features of its CM diagram (see Gratton et al. 2012 for
  further discussion).}  (see Sect.~5 for the case of NGC~2808).
Quantitative simulations of the HB in this and other clusters might
help to determine which of the models presented is the best. While
these simulations are outside the scope of this paper, we point out
that the models presented can potentially explain a variety of
different HB morphologies and can be tested by spectroscopic
observations similar to those by \cite{marino2011,
  villanova2012}\footnote{  We comment, in passing, on the case of
  NGC~6397, a cluster having a short HB and a thin MS, both indicative
  of scarce star--to--star differences in helium
  \citep{dicriscienzo63972010} but a significant spread in
  sodium. This latter was assumed by \citep{carretta2010} to
  correspond to the mixture of FG plus SG stars, with $\sim$30\% of
  stars belonging to the FG. We can not model NGC~6397 with the
  present yields, as it has a much smaller metallicity, but we remark
  that it can be reanalysed in terms of a model similar to the SG of
  model M~4--5 in Fig.~\ref{figuranuova}, {\it assuming that the FG of
    NGC~6397 has been fully lost, as suggested by
    \cite{dantonacaloi2008}.}}.

We summarize these results by concluding that:
\begin{enumerate}
\item The mild O--Na anticorrelation and the small -- if any --
  variation in helium shown by a class of GCs can be due to pollution
  of super--AGB stars into the re--accreted pristine gas forming the
  second generation stars.  The needed ingredient is that the pristine
  matter falls back to the cluster core very soon after it has been
  totally removed from the cluster, during the FG SN~II epoch.
\item Another important outcome of the simulations including the
  super--AGB ejecta is that the SG formation phase can have a short
  duration ($\sim$ 10-20 Myr) and does not need to extend up to $\sim$
  100 Myr (as found by D2010). The models characterized by a shorter
  SG formation epoch, on the other hand, require a larger star
  formation efficiency\footnote{The hydrodynamical simulations by
    \cite{dercole2008} proved to be independent of the efficiency of
    star formation ---see their Appendix A. This remains true {\it
      when the timescale of star formation is long enough that all the
      gas is consumed}. For models with shorter SG formation timescale
    the role of $\nu$\ becomes critical.}.  Models with a short SG
  formation phase {\it can be compatible with a second phase of SN~II
    explosions due to the evolution of SG massive stars}, thus
  relaxing any assumption of anomalous IMF.
\item As an aside of the previous point, we stress that a shorter SG
  formation implies that the mass range of the AGB progenitors
  providing polluted ejecta for the SG formation is narrower and,
  therefore, that these models require a more massive FG cluster to be
  able to produce the observed amount of SG stars. Models with a short
  SG formation phase typically require an initial FG mass 3--4 times
  larger than that needed in models with a more extended SG formation
  phase.
\end{enumerate}

\begin{figure}
\vskip -25pt
\centering{\includegraphics[width=8cm]{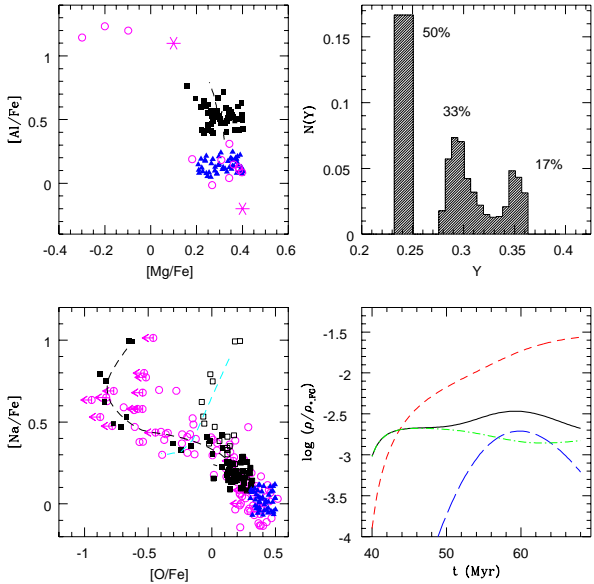}}
\vskip -50pt
\caption{Results for the model NGC~2808-0. Data for giants (open
  circles) are taken from Carretta et al. (2009a, 2009b). The two
  turnoff stars examined by Bragagliaet al. (2010) are shown in the
  Mg--Al panel with star symbols. The open squares in the
  O--Na panel represent the location of stars formed from ejecta with
  Y$>$0.34 in absence of deep-mixing (e.g. for the blue MS turnoff
  stars). The remaining symbols are as in Fig. \ref{f4}.  The top
  right panel shows the histogram of the helium distribution, with
  relative number fractions in evidence.}
\label{f0} 
\end{figure}

\begin{figure}
\vskip -25pt
 \centering{\includegraphics[width=8cm]{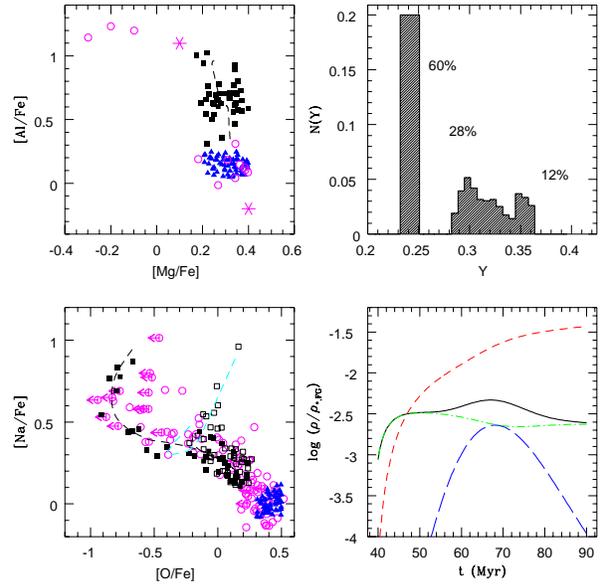}}
\vskip -50pt
\caption{As Fig. \ref{f0}, but for model NGC~2808--1.}
\label{f2} 
\end{figure}

\begin{figure}
\vskip -25pt
 \centering{\includegraphics[width=8cm]{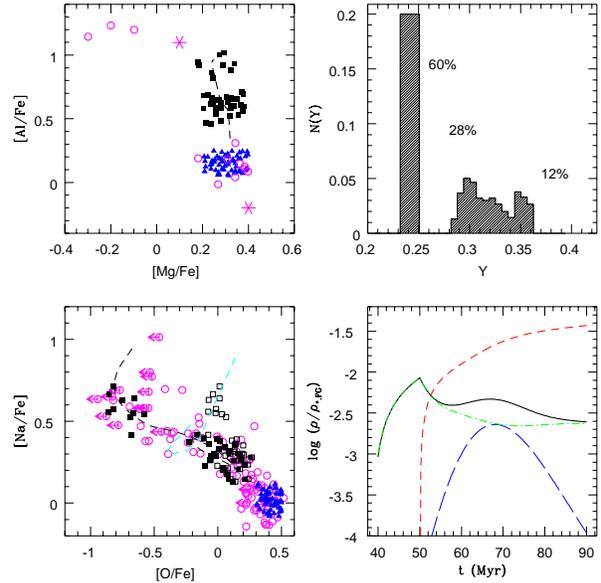}}
\vskip -50pt
\caption{Results for the model NGC~2808--2 . The parameters are the
  same as to those of model NGC 2808-1 in Fig. \ref{f2}, but the
  super--AGB ejecta are assumed to accumulate without forming stars
  until $t=50$~Myr, as can be seen from the bottom right panel
  (symbols as in Fig. \ref{f0}). }
\label{f3} 
\end{figure}

\section{Results: NGC~2808}
\label{sec:n2808}

{ We model NGC~2808 by taking it as the prototype of clusters with
  an extended Na--O anticorrelations, along the lines described by
  D'Ercole et al. (2008) and in D2010. This means that we do not
  attempt to model the recent observational details in the analysis of
  the red HB stars, shown by Gratton et al. (2011), and consider only
  the formation of two main populations: an extreme one, born from
  super--AGB undiluted ejecta, and a milder one, born after dilution
  with the re--accreted pristine gas.}

\subsection{Continuous star formation}
\label{subsec:cont2808}

Figure \ref{f0} shows the model NGC 2808-0 which has a set of
parameters very similar to that for the model of NGC 2808 illustrated
by D2010.  After the phase of formation from super--AGB ejecta, the
peak value of re--accretion occurs at \taupeak=58~Myr (cf. the bottom
right panel). The epoch from 38.9 to $\simgt\sim$50~Myr corresponds to
the formation of the Extreme population, with Y$\sim$0.35. This
accounts for the presence of the blue main sequence. The following
dilution phase leads to the formation of a less extreme SG
(Y$\sim$0.3) that populates the intermediate main sequence of this
cluster.

The top left panel of figure \ref{f0} shows the Mg and Al
predictions. Using the new yields for Mg and Al listed in Table~1 {\it
  does not allow the fit of the three Mg--poor giants
  ($\delta$[Mg/Fe]$\sim$--0.6) observed by Carretta et
  al. 2009b}. \cite{vcd2011} deal extensively with the Mg--Al problem
in the models, and show that small adjustments to the cross sections
(concerning the Mg isotopes) may well reproduce the [Mg/Fe]--[Al/Fe]
value of the blue MS stars observed by \cite{bragaglia2010} and shown
in Fig.~\ref{f0} with the star symbol. { In particular, an increase by a factor two of the
cross section $^{25}$Mg(p,$\gamma$)$^{26}$Al shifts the 6\Msun\ yield in the region
occupied by the M~13 giants and by the MS star of NGC~2808. Such an increase is also 
consistent with recent new measurements of the reaction rate \citep{luna}.}

The difference in Mg between the MS stars and the giants can not be
attributed to further dilution in the red giants, as magnesium is not
touched in the red giant evolution of low mass stars, not even in the
interior \citep{dv2007}.  { Figure 5 in \cite{vcd2011} shows the
  Mg--Al data in the literature for the clusters NGC~6752, M~15, NGC
  2808 and M~13, and the results of different modeling of the yields
  in the chemistry evolution of a 6\Msun\ star as adopted in this
  paper.  Apart from two giants in M~15, a cluster whose metal
  abundance is much smaller than the others, no star has Mg abundances
  as low as those in the three UVES giants of NGC ~2808, that are not
  compatible with the models. As these three stars constitute a
  problem for any other pollution model, and can not be explained in
  the deep extra mixing framework as well, we defer the study of this
  problem to a future investigation.}

Let us now consider the [Na/Fe]-[O/Fe] diagram (bottom left panel of
Fig. \ref{f0}). While in the case of M 4-0 the new simulation is very
similar to that presented in D2010 for the absence of an Extreme
population, here the new super--AGB sodium yields {\it are able to
  account for the high sodium values of the stars with upper limit
  only available for oxygen} (full squares); this results ensues from
the deep-mixing hypotesis, and the subsequent reduction of the oxygen
values given by the recipe described in Sect.~\ref{O16}.  The open
squares represent the values of oxygen that should be found in turnoff
stars for the stars with Y$>$0.34, as in these stars deep-mixing has
not yet taken place.  These stars, as well as blue MS stars, should be
characterized by oxygen abundances larger than those of giant stars
and fall along the Na-O curve determined by the new yields and shown
in Fig. \ref{f1}, a prediction that can be falsified by spectroscopic
observations.
 
The top right panel of Fig. \ref{f0} shows the histogram of the
distribution of helium abundances.  The number ratio of the
Intermediate and Extreme populations of the SG depends on the initial
mass function and on the timing and extent of the dilution with
pristine gas, and thus this is a powerful constraint for the
successful models.  The observed number ratios of the three
populations differing in helium content in NGC~2808 is constrained by
two different sets of observations:
\begin{enumerate}
\item the number counts of the HB stars \citep{bedin2000}. These
  numbers were the basis of the interpretation of the HB in terms of
  multiple populations with different helium content \citep{dc2004,
    dantona2005}, providing fractions of 50\%, 35\% and 15\%,
  respectively, for the groups of standard helium (red clump stars),
  intermediate helium (blue luminous HB, named EBT1) and high helium
  (sum of the EBT2 and EBT3 extensions of the blue HB).
\item the number counts in the triple MS. {\cite{milone2011b} recently
    re--analyzed the data by \cite{piotto2007} in order to derive the
    mass functions of the three populations. According to this
    analysis, the fractions of standard (red MS), intermediate (middle
    MS) and high helium (blue MS) stars are respectively 62\%, 24\%
    and 14\%. Therefore, while the fraction of FG stars changes from
    $\sim$50 to $\sim$60\%, the proportion of Intermediate and Extreme
    population are very similar in the two analyses: the Extreme
    population is $\sim$35\% of the whole SG. }
\end{enumerate}

In the present models for NGC~2808 we set $x=0.4-0.5$, that is, we
assume that the fraction of FG stars, with standard helium content
Y=0.24, is 60-50\%, and set the other input parameters in order to
obtain a reasonable result for the number ratio of the Extreme and
Intermediate SG.

From the above discussion it is clear that, contrary to the case of M
4, there is less freedom in the choice of the parameters because of
the presence of the Extreme population and the need to reproduce the
right proportion among the Extreme, the Intermediate and the
Primordial populations (cf. the three peaks in the He histogram in
Fig. \ref{f0}). Some degree of flexibility is still present though.
Figure \ref{f2} shows another model for NGC~2808, whose input is model
NGC~2808-1 in Table 2.  In the left panels we see that, as for the
model NGC 2808-0, when dilution with pristine matter sets in, the
model trajectories (dashed line) tend towards the FG abundances, but,
during the last 10--15 Myr, star formation occurs again in mostly
undiluted AGB ejecta, and the abundances revert back to lower O and
larger Na values. In fact, this last phase of star formation is
necessary to fill the gap in O, Na values between the formation of the
Extreme population (with very large Y, corresponding to the blue MS)
and the Intermediate population with strong dilution.  The bump in the
helium distribution between the peaks at Y=0.3 and Y=0.35 (top right
panel) forms during this same stage as well.  And also the stars with
very large [Al/Fe] in the simulation (upper left panel) form in this
last epoch of star formation. This can be understood by looking at the
abundances listed in Table~1, showing that the maximum abundance in Al
and maximum depletion in Mg are reached with the evolution of the
5.5\msun (at about 85 Myr).

\subsection{Star formation after gas accumulation}
\label{subsec:accum2808}

We now focus on Fig.~\ref{f3} that shows the results of the model
NGC~2808--2, having the same input parameters of model NGC 2808--1,
but dealing with a different hypothesis for the star formation: here
we assume that SG formation does not start immediately after the end
of the SN~II epoch, but that gas accumulates from 40 to 50~Myr, and
then a star formation burst occurs.

The results are very similar, but the Extreme population does not
reach the high sodium abundances obtained in the model NGC~2808--1;
this is a consequence of the fact that, before the SG formation
starts, the ejecta of the stars between $8\msun$ and $\sim$7.3\msun
mix, so that the highest abundances are lowered.

\subsection{Models with interruption of star formation due to SG SN~II explosions}
\label{sec:snsg}

\begin{figure}
\vskip -25pt
  \centering{\includegraphics[width=8cm]{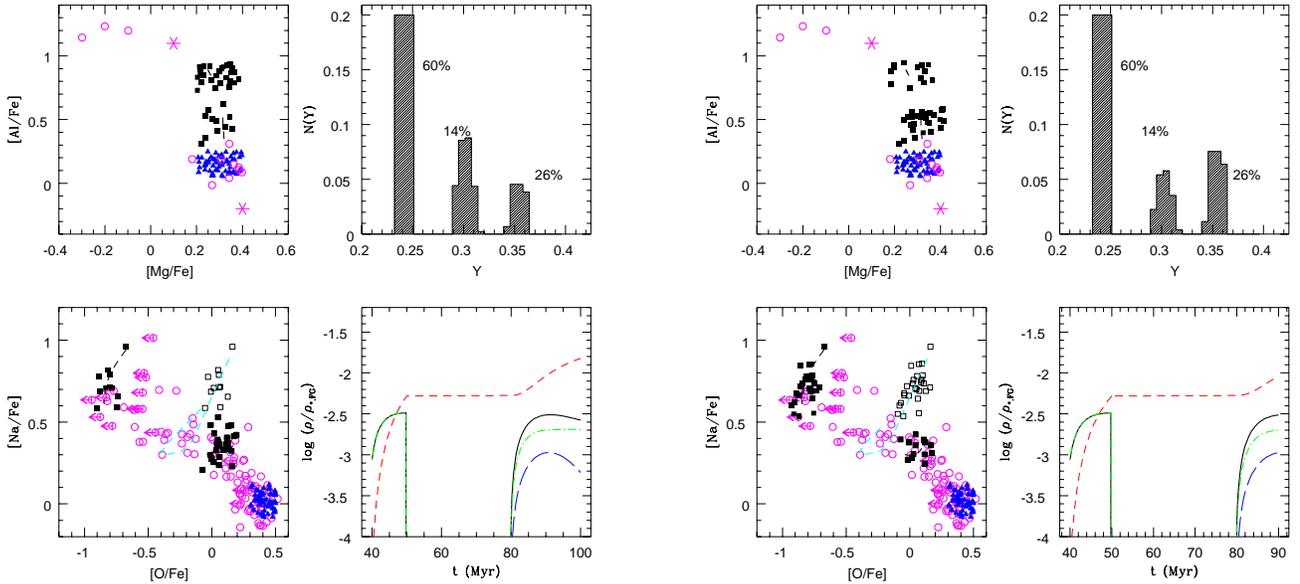}}
\vskip -50pt
\caption{Case NGC~2808-3. Star formation is interrupted for 30~Myr,
  but the second stage of star formation lasts for 20~Myr (symbols as
  in Fig. \ref{f0}).}
\label{f9} 
\end{figure}
\begin{figure}
\vskip -25pt
 \centering{\includegraphics[width=8cm]{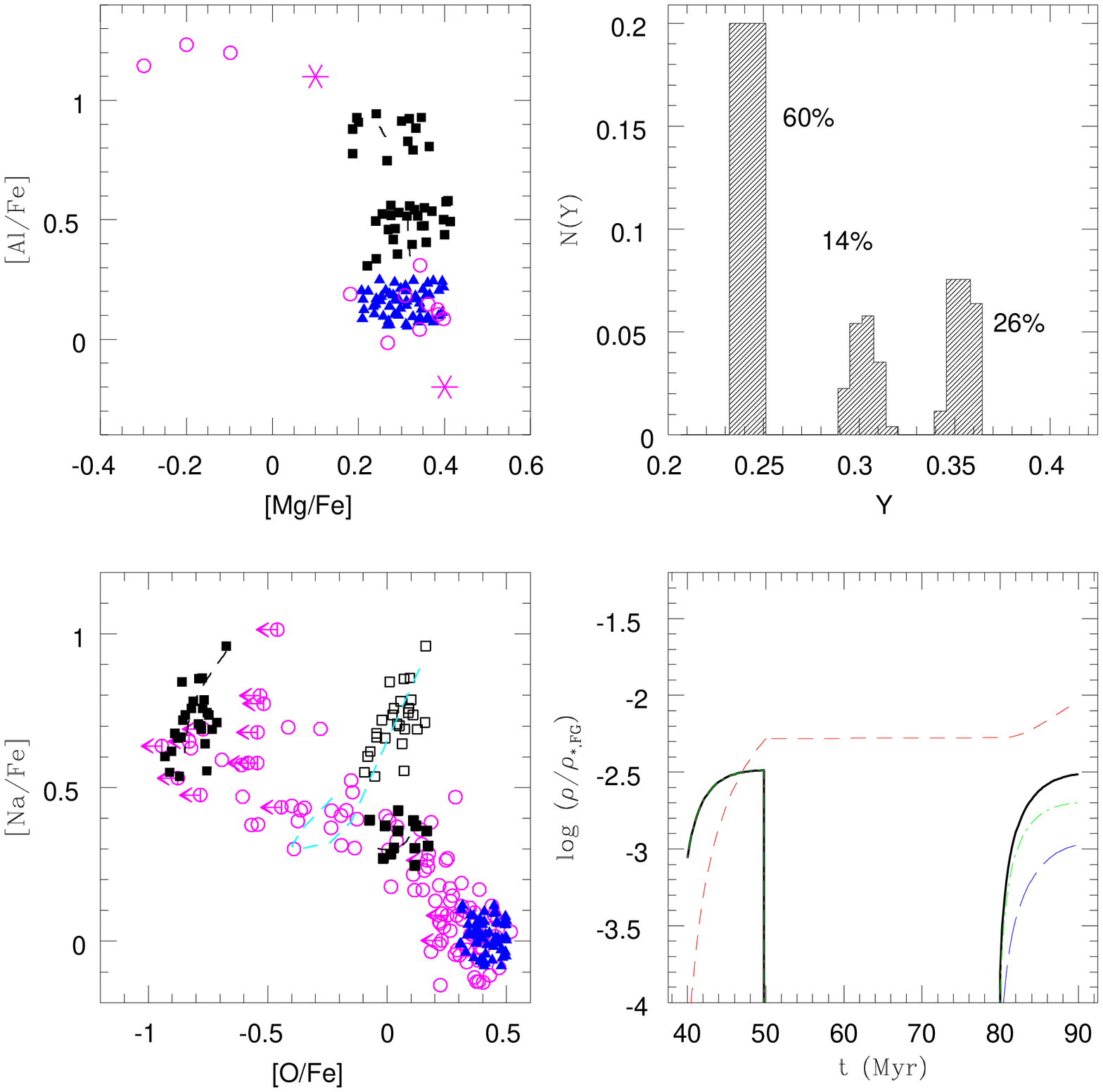}}
\vskip -50pt
 \centering{\includegraphics[width=8cm]{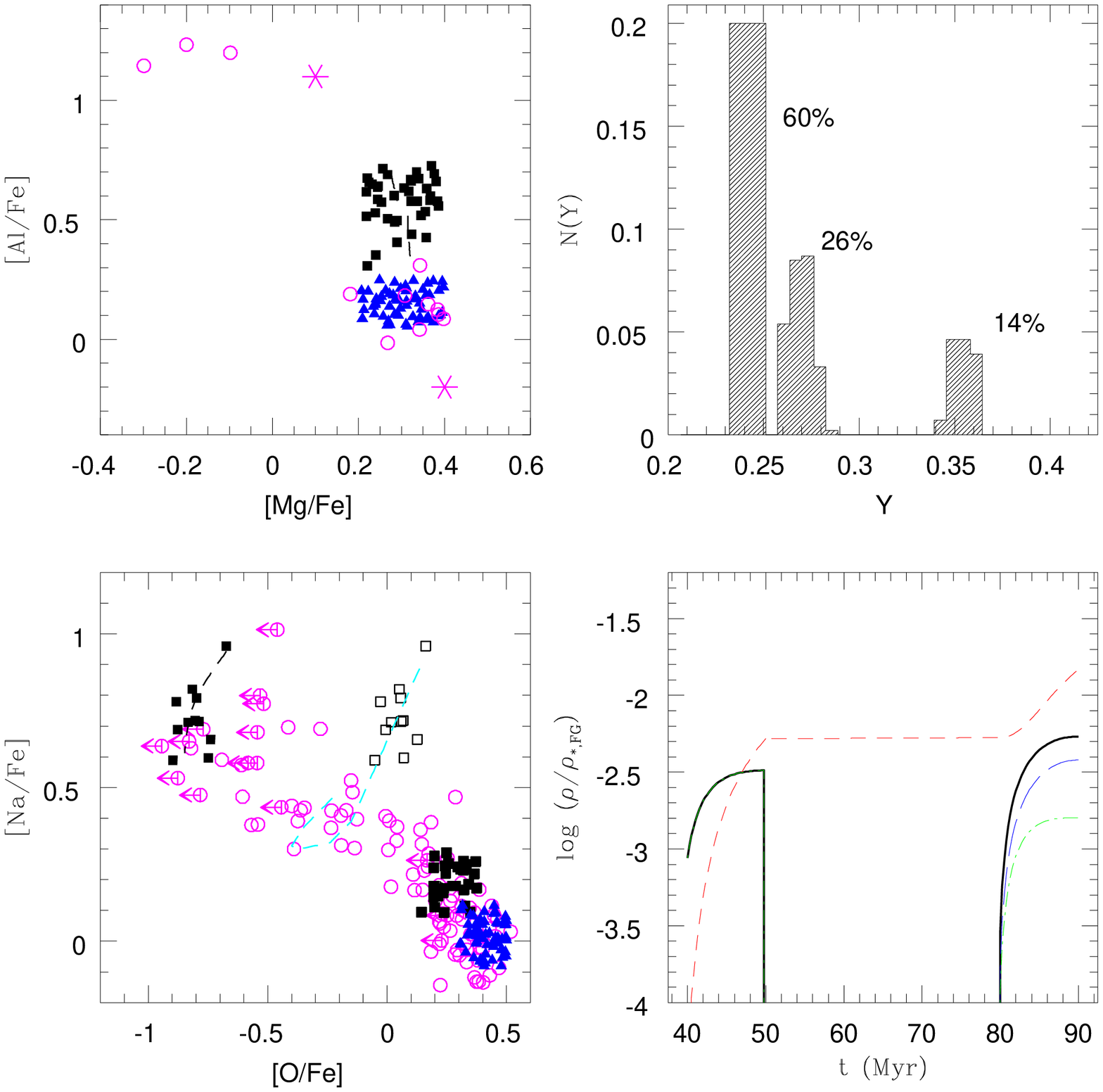}}
\vskip -50pt
\caption{Case NGC~2808-4 (top) and 5 (bottom). Using the input parameters of
  Fig.~\ref{f9}, but limiting the second star formation phase to
  10~Myr, the Intermediate population is very scarce (top
  panel). Increasing the infall of pristine matter, the O--Na data
  come very close to the FG values (bottom panel) (symbols as in
  Fig. \ref{f0}).}
\label{f10} 
\end{figure}

As discussed in Sect.~\ref{sec:results}, an interesting implication of
clusters like M 4, with a short [O/Fe]-[Na/Fe] anticorrelation, is
that thay can form the bulk of the SG population in $~10$ Myr, or even
less; such a timescale is compatible with the survival of a cooling
flow in the cluster core, previous to the explosion of the Type II SN
{\it of the second generation}\footnote{Although the most massive
  stars may take only 5~Myr to accomplish their complete evolution, it
  is possible that only stars below a mass evolving in $\geq$10~Myr
  produce a SN~II explosion, the more massive stars may end their life
  in the less energetic formation of a black hole \citep{smartt2009}},
and thus there would be no reason to limit the IMF of the SG to the
stars that do not explode as SN II.

Understanding whether the formation of two distinct SG populations
(such as those found in NGC~2808) is compatible with star formation
bursts lasting $\sim$10~Myr is not straightforward. We have tried to
model in detail the three populations of NGC~2808 by making the
assumption that, $\sim$10~Myr after the Extreme population began to be
formed in the cluster, a new phase of SN~II begins. This will hamper
star formation until the end of the SN~II phase.

In our models, this phase lasts for $\simlt$38~Myr (Table 1). We can
have a second burst of star formation from the AGB ejecta starting at
$\sim$90~Myr. We have discussed in Sect.~\ref{sec:yields} that the
time hiatus between the two bursts may be reduced to $\sim$30~Myr for
models employing a smaller core--overshooting. Thus we employ in the
following $\sim$30~Myr as total duration of the SG SN~II
epoch\footnote{Notice that the timing of each mass for our
  simulations, given in Table~1, is still adopted in the same way in
  these models.}.

The main difficulty encountered in models including SN~II explosions
from the Extreme SG is to reproduce at the same time the extension and
distribution of stars along the O--Na anticorrelation and the helium
distribution function.
\begin{itemize}
\item Case NGC~2808-3 (Fig.~\ref{f9}): In this model we stop the SG
  formation from pure ejecta at 50~Myr, allow for a $\simgt$30~Myr
  period without star formation, and then resume star formation from
  the AGB ejecta mixed with pristine gas, with a choice of
  \taupeak=87~Myr. The clumps in the distribution of stars in the
  O--Na plane are probably too separated to be consistent with the
  observations.  Moreover, in order to obtain a reasonable number
  ratio between the Extreme stars (Y$>$0.34) and the Intermediate
  population (Y$\sim$0.30), we are forced to extend the star formation
  of this latter population to $\sim$20 Myr, much longer than allowed
  by the beginning of the new (third) SN II epoch.
\item Case NGC~2808-4 (Fig.~\ref{f10}, top panel): this model is
  identical to NGC~2808-3, but we limit the formation of the
  Intermediate population to 90~Myr: as discussed above, the
  Intermediate population is a bare 11\%, too small to reproduce the
  number ratio of the Extreme-- and Intermediate--helium MS.
\item (Case NGC~2808-5, Fig.~\ref{f10}, bottom panel): we increase the
  relative number of Intermediate population stars by increasing the
  dilution. The second peak of helium content occurs at Y$\sim$0.28,
  and the number ratio of Extreme and Intermediate population becomes
  reasonable, but the O--Na values become much too close to the FG
  values, due to the strong dilution (a problem discussed in
  \cite{dercole2011} in a different context).
\end{itemize}

We find then that these simulations do not reproduce the observations
so well as models NGC~2808-1 and 2. Nevertheless, the variety of
observational results for different clusters may have their
explanation in the variety of relevant parameters. For instance, model
NGC~2808-4 (Fig.~\ref{f10} top panel) could be relevant to explain the
horizontal branch stellar distribution of the cluster NGC~2419
\citep{dicriscienzo2011}.

\section{Conclusions}
\label{sec:discussion}
In this paper we have explored the possible role of super-AGB in the
formation of the observed abundance patterns in multiple-population
globular clusters.  Our models are based on the set of yields
calculated by \cite{vd2009,vd2011}.  The results of our investigation
show that the new super--AGB yields broaden the range of possible
formation histories leading to the observed abundance patterns.
Specifically:
\begin{itemize}
\item we have built models for M~4, a cluster representative of
  numerous other GCs characterized by a shorter Na--O
  anticorrelation. For this cluster, we show that the observed
  abundance pattern can be reproduced also by models with a short SG
  formation phase ($\sim$ 10-20~Myr) involving only super-AGB ejecta
  and pristine gas. We confirm that models presented in our previous
  work (D'Ercole et al. 2010) which are characterized by a more
  extended SG formation epoch involving super--AGB and AGB ejecta
  (along with pristine gas) can also reproduce the observed abundance
  patterns. While models with a shorter SG formation epoch can include
  the effect of SG SN~II, the narrower range of AGB progenitors
  providing gas for SG formation implies that they require a larger FG
  initial mass (by a factor 3-4 compared to models with a more
  extended SG formation phase).
\item In order to account for the O-poor Na-rich stars of NGC 2808 we
  assumed the presence of deep-mixing in SG giants forming in a gas
  with helium abundance Y$>0.34$, which significantly reduces the
  atmospheric oxygen content, while preserving the sodium abundance.
  The use of the super--AGB yields does not change the main
  requirements on SG formation history of models for the more complex
  clusters like NGC~2808. In these clusters, an Extreme population
  with very large helium content form from the pure ejecta of
  super--AGB stars, followed by formation of an Intermediate
  population by dilution of massive AGB ejecta with pristine gas.  We
  also attempted to model the three populations of NGC~2808 allowing
  for a long hiatus between the formation of the Extreme and
  Intermediate population (possibly due to SN~II explosions of the
  Extreme populations), but our attempts to model each star formation
  epoch in a period as short as $\sim$10~Myr were not
  satisfactory. Nevertheless, further study is probably needed to
  settle this issue.
\end{itemize}

We finally want to point out that one-zone models such as those
presented here (and in D2010), while very useful to understand rather
easily and very quickly the role of different ingredients in shaping
the GC chemical evolution (AGB ejecta, amount of pristine gas,
duration of the system evolution, etc.), cannot of course capture the
complexity and variety of effects connected with the gas
hydrodynamics. As a simple example, we mention that, during the
cooling flow evolution of the GC gas \citep[c.f.][]{dercole2008}, most
of it cumulates in the centre of the cluster which therefore at each
time hosts not only the AGB ejecta delivered ``in situ'', but also
that delivered at earlier times by FG stars located at the GC
outskirts. It is thus expected that SG stars with different chemical
properties can form in different places at the same time, and not only
at different times as in one-zone models. For this reason we plan to
work out hydrodynamic models similar to those of \citet{dercole2008},
but taking into account the detailed chemistry as in the present
models.

 \section{Acknowledgments} 
 We thank Raffaele Gratton for useful comments that helped us to
 clarify and further expand some of the points presented in the paper.
 This work has been supported through PRIN INAF 2009 "Formation and
 Early Evolution of Massive Star Cluster".  EV was supported in part
 by grant NASA-NNX10AD86G.

\label{lastpage}

\end{document}